\documentclass[lettersize,journal]{IEEEtran}
\usepackage{amsmath,amsfonts}
\usepackage{url}
\usepackage{algorithmic}
\usepackage{algorithm}
\usepackage{array}
\usepackage[caption=false,font=normalsize,labelfont=sf,textfont=sf]{subfig}
\usepackage{caption}
\usepackage{cite}
\usepackage{graphicx} 
\usepackage{stfloats}
\usepackage{subcaption}
\usepackage{textcomp}
\usepackage{verbatim}
\usepackage{xcolor}
\usepackage{hyperref}
\usepackage{svg}
\usepackage[normalem]{ulem}

\begin{document}

\title{Direct-to-Cell: A First Look into Starlink's Direct Satellite-to-Device Radio Access Network through Crowdsourced Measurements}

\author{
\IEEEauthorblockN{Jorge Garcia-Cabeza\IEEEauthorrefmark{1}\IEEEauthorrefmark{2}, Javier Albert-Smet\IEEEauthorrefmark{1}, Zoraida Frias\IEEEauthorrefmark{1},
\\Luis Mendo\IEEEauthorrefmark{1}, Santiago Andrés Azcoitia\IEEEauthorrefmark{1}, and Eduardo Yraola\IEEEauthorrefmark{2}}
\\
\IEEEauthorblockA{\IEEEauthorrefmark{1}\textit{Universidad Politécnica de Madrid}. Madrid, Spain\\
Email: \{javier.albert.smet,\,zoraida.frias,\,luis.mendo,\,santiago.andres\}@upm.es} 
\\
\IEEEauthorblockA{\IEEEauthorrefmark{2}\textit{Weplan Analytics}. Madrid, Spain\\ Email: \{jorge.garcia,\,eduardo.yraola\}@weplananalytics.com}

\thanks{\IEEEauthorrefmark{3} \textbf{Corresponding author: Zoraida Frias}, zoraida.frias@upm.es \\ This work has been partially supported by Weplan Analytics through the CROWD-X project and by UPM's \textit{Doctores Emergentes} program (24-ED3B0Y-100-JOQZ29), funded by the Region of Madrid.}

}

\markboth
{This work has been submitted to the IEEE for possible publication. Copyright may be transferred without notice}{}

\maketitle

\begin{abstract}
Low Earth Orbit (LEO) satellite mega-constellations have emerged as a viable access solution for broadband connectivity in underserved areas. In 2024, Starlink, in partnership with T-Mobile, began beta testing an SMS-only Supplemental Coverage from Space (SCS) service. This marks the first large-scale deployment of Direct Satellite-to-Device (DS2D) communications, allowing unmodified smartphones to connect directly to spaceborne base stations. This paper presents the first measurement study of deployed DS2D technologies. Using crowdsourced mobile network data from the U.S.~between October 2024 and July 2025, we provide evidence-based insights into the capabilities, limitations, and future evolution of DS2D technologies for extending mobile connectivity. We find a strong correlation between the number of satellites deployed, the number of unique cell identifiers measured, and the volume of measurements, concentrated in accessible areas with poor terrestrial network coverage, such as national parks and sparsely populated counties. Stable physical-layer measurements were observed throughout the period, with a $\mathbf{24}$-dB lower median RSRP and a $\mathbf{3}$-dB higher RSRQ compared to terrestrial networks, reflecting the SMS-only usage of the DS2D network during this period. Based on the SINR measurements collected, we estimate the expected performance of the announced DS2D mobile data service to be around $\mathbf{3}$ Mbps per beam in outdoor conditions. We also discuss strategies to expand this capacity up to $\mathbf{18}$ Mbps in the future, depending on key regulatory and business decisions, including allowable out-of-band emissions, permitted number of satellites, and availability of spectrum and orbital resources. 
\end{abstract}


\section{Introduction}
\IEEEPARstart{O}{ver} the past decade, mobile networks have become an essential means of accessing the Internet for billions of users worldwide. Despite a remarkable growth in mobile network coverage, significant portions of the global population remain underserved by terrestrial cellular infrastructure, particularly in remote, rural, and sparsely populated areas.

In recent years, particularly from 2020, Low Earth Orbit (LEO) mega-constellations have become a viable and increasingly widespread Non-Terrestrial Network (NTN) Internet access solution for fixed broadband services. Networks like SpaceX's Starlink, Eutelsat's OneWeb or Amazon's Kuiper are successfully addressing longstanding limitations of traditional geostationary satellite communications. By operating at significantly lower altitudes ($500$--$1200$ km), these constellations offer latency levels comparable to those of terrestrial networks (typically tens of milliseconds vs.~the hundreds associated with geostationary satellites), supporting real-time applications such as video conferencing and online gaming~\cite{Mohan24Local}.

Following technological developments in fixed broadband connectivity, satellite-based cellular communications, known as Direct Satellite-to-Device (DS2D), have become a promising solution to provide seamless connectivity directly to standard unmodified smartphones. Starlink, the satellite communications division of SpaceX, has emerged as the global leader in DS2D communications by rapidly deploying a non-3GPP-NTN-compliant solution that leverages terrestrial International Mobile Telecommunications (IMT) spectrum. In partnership with T-Mobile and using its spectrum, Starlink has been the first company to launch large-scale beta testing of a Supplemental Coverage from Space (SCS) service. The system is initially limited to Short Message Service (SMS), with support for voice and data services planned by late 2025. 

In this article, we examine Starlink's SCS services and characterize its DS2D network---referred to by Starlink as \textit{Direct-to-Cell}. Using a large-scale dataset provided by the company Weplan Analytics comprising millions of crowdsourced measurements collected from October 2024 to July 2025, our analysis provides valuable insights into the implications of DS2D network design for mobile network planning and spectrum policy. The main contributions of this work are:

\begin{itemize}
    \item The first measurement-based study of a commercially operating DS2D network, encompassing both spatial and temporal dimensions. 
    \item A large-scale empirical evaluation of physical-layer Radio Access Network (RAN) parameters of Starlink's DS2D network, namely Reference Signal Received Power (RSRP), Reference Signal Received Quality (RSRQ) and Signal-to-Interference-plus-Noise Ratio (SINR).
    \item An assessment of the projected performance of Starlink's future SCS data services and strategies for network capacity expansion.
\end{itemize}

The remainder of this paper is structured as follows. Section~\ref{Section:Background} provides background on the evolution of LEO satellite networks and summarizes related research. Section~\ref{Section:Methodology} describes the dataset and our analytical approach. Section~\ref{Section:Results} presents and discusses the results. Finally, Section~\ref{Section:Conclusions} offers conclusions and suggests future research.

\section{Background} \label{Section:Background}
\subsection{Technological Evolution and Emerging Ecosystem}

LEO satellite communications have progressed considerably in recent decades, driven by satellite miniaturization, improved launch capabilities, and the growing demand for ubiquitous connectivity. Early LEO constellations in the late 1990s were primarily designed to deliver voice and low-speed data services to remote and underserved areas. A new phase in LEO communications emerged in the late 2010s with the advent of mega-constellations~\cite{Su2019BroadbandTechnologies}. Initiatives such as SpaceX’s Starlink and OneWeb began deploying extensive LEO satellite networks to provide broadband Internet access globally.

DS2D has been further enabled by significant progress in electronics, embedded systems, digital signal processing, and beamforming technologies, among others \cite{Starlink2024DS2D}. Specifically, DS2D connectivity relies on the satellites being equipped with advanced phased array antennas, highly sensitive radio receivers, and high-power transmitters. Moreover, compensation for Doppler shift and delay is essential to overcome the physical constraints of mobile-to-satellite links, and software-defined radio (SDR) is necessary to adapt the system to varying frequencies and technologies.

3GPP specifications have recently incorporated NTN in Release 17. Protocols have been enhanced to address the distinctive characteristics of satellite-based communications, such as the larger propagation distance between user terminals and satellites, which impacts procedures including hybrid automatic repeat request (HARQ) and random access. In addition, system information blocks have been extended to convey satellite-specific parameters, such as orbital position. Subsequent releases introduce further enhancements, focusing on improved coverage (potentially extending to indoor environments), support for neighbor cell measurements, and reduced battery consumption.

Starlink has pioneered DS2D deployments with a strategy focused on a dense LEO constellation at the lowest feasible orbital altitudes (currently primarily $\sim 550$ km and selectively $\sim 350$ km subject to coordination with NASA). In contrast, its main competitor, AST SpaceMobile, relies on a less dense constellation at higher altitudes ($\sim 700$ km), supported by very large antenna arrays. Although both operators have so far relied primarily on IMT spectrum obtained through agreements with mobile network operators in different countries, they are now pursuing their own terrestrial and satellite spectrum. Other actors in the emerging DS2D ecosystem include Lynk Global, Skylo, and Viasat-Inmarsat. Key use cases comprise ubiquitous connectivity for IoT, supplementary coverage for terrestrial mobile networks, SOS alerts, emergency communications, and maritime and aeronautical connectivity.

\subsection{Starlink's Direct Satellite-to-Device (DS2D) Approach}

Starlink has reshaped the satellite telecommunication sector by deploying over $7\,000$ satellites since its first launches in 2019. The bent-pipe architecture initially adopted \cite{Ma2023NetworkUsersLocal} enabled rapid deployment using simpler satellite designs at the cost of deploying a dense network of ground stations to ensure continuous coverage. From 2022 onward, Starlink progressively introduced laser inter-satellite links (ISL), fully adopted in its second-generation (``Gen2'') fleet, enabling onboard signal processing and space-based routing. This transition from a ground-dependent relay to a space-based mesh brings expanded coverage, improved resilience, and enhanced end-to-end performance \cite{Mohan24Local}.

Rather than awaiting the completion of 3GPP standardization of NTN, Starlink pursued an early deployment strategy by integrating conventional LTE eNodeB payloads into its satellites, complemented with proprietary adaptations for NTN operation. This allows smartphones to connect natively using terrestrial mobile standards, while traffic is handled either through the company’s ground station network or, when necessary, routed across the constellation via ISLs \cite{Chaudhry2021LaserAnalysis}.

SpaceX announced the completion of its first DS2D constellation in December 2024, consisting of approximately 400 satellites---barely $6\%$ of their deployed fleet. In addition to the SCS service provided in the U.S.~through the partnership with T-Mobile, the constellation also offers SCS connectivity in New Zealand and Australia via a partnership with operators and is under testing with operators in several other countries.

In November 2024, the FCC granted partial approval for Starlink's DS2D operations to initiate the SMS-only beta testing, but imposed several conditions to prevent harmful interference with terrestrial and satellite incumbents, including the requirement to stop transmission immediately upon any substantiated claim of interference, limitations on power flux density levels for aggregate out-of-band emissions (OOBE), and mandatory coordination procedures with incumbents.\footnote{See FCC's DA 24-1193.} 

Subsequently, in March 2025, the FCC approved\footnote{See FCC's DA 25-197.} a conditional waiver allowing Starlink to increase
\label{part: 10-dB increase}
its OOBE by $9.4$ dB, despite formal objections from terrestrial operators, including AT\&T and Verizon, and EchoStar. The waiver stipulated that Starlink must continue to adhere to strict OOBE limits and engage in active interference resolution with affected parties.

In the U.S., Starlink’s DS2D beta operations have relied exclusively on the PCS G Block, a $2 \times 5$ MHz channel at $1910$--$1915$ MHz (uplink) and $1990$--$1995$ MHz (downlink), which is part of the IMT spectrum licensed to T-Mobile across the contiguous United States.\footnote{See FCC's IBFS File No.: SAT-LOA-20210511-00064.} In September 2025, having concluded the beta phase, Starlink acquired Mobile Satellite Service (MSS) spectrum rights from Echostar in the AWS-4 band ($2000$--$2020$ MHz uplink, $2180$--$2200$ MHz downlink, $2 \times 20$ MHz) and IMT PCS H Block ($1915$--$1920$ MHz uplink, $1995$--$2000$ MHz downlink, $2 \times 5$ MHz). These additional allocations significantly expand the spectrum available for its Direct-to-Cell operations, particularly since the AWS-4 band overlaps with the 5G NR NTN band n252.
\label{AWS-4, PCS H}

\subsection{Related Work}
Mirroring the rapid pace of innovation of LEO satellite communication technologies, a growing body of research has emerged from multiple perspectives to explore the design, deployment, and operation of these systems. Early studies focused primarily on the technologies that underpin LEO broadband services, addressing alternative network architectures (e.g., bent-pipe versus ISL designs), hybrid beamforming strategies, interference coordination mechanisms, and resource management challenges \cite{Su2019BroadbandTechnologies}. Although much of this work was initially centered on fixed satellite broadband services, recent attention has shifted toward DS2D communications \cite{Pasandi2024AProspectsLocal}, which is reflected in dedicated research and publication initiatives.\footnote{See the Special Issue \textit{Direct Satellite-to-Device Communications: Technologies, Connectivity, and Spectrum Management} of the IEEE Communications Magazine (2025).} Alongside theoretical and architectural advances, new simulation tools, such as Hypatia and StarPerf, have been developed to support realistic modeling, design, and performance evaluation of LEO satellite networks under operational conditions.

Following initial deployments and the establishment of operational networks, a growing number of measurement studies have emerged, focusing on Starlink’s fixed satellite broadband services ~\cite{Mohan24Local, MichelAPerformanceLocal, Laniewski2025MeasuringLook, Laniewski2024StarlinkEuropeLocal}. These studies have primarily evaluated connectivity delivered through dedicated fixed user terminals (i.e., satellite ``dishes''), with particular attention to metrics such as throughput and latency, a major historical limitation of traditional geostationary satellite systems.

Despite growing research on LEO satellite networks, to the best of our knowledge, there are no empirical evaluations of commercial DS2D deployments. This paper presents the first measurement-based analysis of a real-world DS2D network, using large-scale crowdsourced mobile data collected during the beta testing phase of Starlink’s service in the U.S.

\section{Methodology}
\label{Section:Methodology}
In this study, we adopt a data-driven, user-centric methodology based on large-scale crowdsourced mobile measurements. Crowdsourced measurements offer a unique vantage point for evaluating cellular network performance, capturing real-world user experience even in sparsely populated areas \cite{Frias2025BuildingChallengesLocal}. The dataset is provided by the crowdsourcing company Weplan Analytics, whose data collection framework is specifically designed to capture and evaluate RAN performance. While the raw crowdsourced measurements are proprietary and cannot be shared, the derived statistics and analytical procedures are fully described in this paper to enable methodological reproducibility. Subsection~\ref{Subsection:Datasources} describes the data sources and the data preprocessing; Subsection~\ref{Subsection:EstimatingSCSUse} presents how we estimate the potential share of SCS services; and Subsection~\ref{Subsection:methodology_performance} details how we calculate the expected performance of DS2D communications based on SINR measurements.

\subsection{Data sources}
\label{Subsection:Datasources}
Our primary data source consists of LTE crowdsourced measurements passively collected from Android user devices between October 2024 and July 2025. All data were anonymized at collection time by the crowdsourcing provider to ensure full compliance with applicable data protection regulations. 
We used the following network variables: (\textit{i}) RAN parameters RSRP, RSRQ, SINR, EARFCN, ECI (E-UTRAN Cell Identifier) as defined in 3GPP specifications~\cite{Johnson12}, and (\textit{ii}) PLMN (Public Land Mobile Network) identifier, comprising Mobile Country Code (MCC) and Mobile Network Code (MNC). We filter with $\mathrm{MCC} = 310$ and $\mathrm{MNC} = 830$ or $210$ for Starlink measurements, and $\mathrm{MNC} = 260$ for T-Mobile's terrestrial network measurements, which we compare with Starlink's.

In addition, we incorporate three complementary data sources that help assess the consistency and reliability of the crowdsourced dataset. First, we compiled a dataset tracking the evolution of the cumulative number of Starlink's DS2D-capable satellites in orbit by aggregating information from official SpaceX mission reports, which we publish openly as supplementary material \cite{rknd-y219-25}. Second, we used the publicly available T-Mobile coverage map to evaluate how well the spatial distribution of DS2D measurements aligns with areas designated as lacking terrestrial coverage. Finally, we used official open-data vector files for U.S.~counties and national parks to quantify the SCS share, as detailed next, and official county-level population data from the U.S.~Census Bureau to study the correlation of SCS share and population density. 

\subsection{Measuring SCS share}
\label{Subsection:EstimatingSCSUse}

To estimate the likelihood of relying on Starlink SCS services in a particular geographic area, we define a spatial indicator called SCS share. This metric quantifies the relative prevalence of Starlink DS2D connectivity in our dataset by computing the ratio of Starlink DS2D measurements to the total number of measurements observed (i.e., Starlink DS2D and T-Mobile terrestrial network measurements across all technologies and frequency bands). It thus indicates the degree to which Starlink serves as the primary mobile coverage provider in a given area. 

\subsection{Estimating DS2D performance} \label{Subsection:methodology_performance}
To assess the performance of future DS2D data services, we use the modified Shannon formula for LTE introduced in \cite{Mogensen07} to estimate the downlink spectral efficiency of the system, $\eta$ (bps/Hz per beam). This formula, for the case of no spatial multiplexing, is $\eta = \min\{s \, a \log_2(1+\textrm{SINR} / b),\, m\}$, where $s$ is a bandwidth efficiency factor, which accounts for all system overhead including guard bands, pilot symbols and signaling channels; $a$, $b$ are fitting coefficients, which model implementation-related losses; and $m$ is a limit imposed by the highest modulation and coding scheme (MCS) that the system can use. The terms $s$ and $m$ can be computed theoretically, whereas $a$, $b$ are adjusted from simulation. We consider a static (no fading) channel model with AWGN and no spatial multiplexing. In these conditions, \cite{Mogensen07} gives the values $s = 0.57$, $a = 0.9$, $b=1.25$. The parameter $m$ depends on the specific technology used by the LTE network. The highest supported MCS can be $64$-, $256$- or $1024$-QAM, with a coding rate close to $1$. As a reference, we use $256$-QAM, for which the best channel-quality indicator (CQI) that the mobile terminal can report corresponds to $7.41$ information bits per symbol \cite[Table 7.2.3-2]{3GPP36213-v17.1.0}. Taking into account the efficiency factor $s$, which already incorporates the effect of the cyclic prefix (as well as other types of overhead), this yields $m=4.22$ bps/Hz. Thus, the expression for spectral efficiency results in: $\eta = \min\{0.51 \log_2(1+\mathrm{SINR}/1.25), 4.22\}$. Following \cite{Mogensen07}, the spectral efficiency is computed for each SINR value, and then averaged over the SINR distribution.

\section{Results}
\label{Section:Results}

This section presents our empirical findings. We start by assessing the relationship between the observed number of cell identifiers and the satellites in orbit in Subsection~\ref{subsection:observed_network_expansion}. Subsection~\ref{subsec:Spatial_Coverage} then examines the spatial availability of Starlink’s SCS service, characterizing the evolution of its geographic footprint, and quantifying its SCS share across different types of areas. Subsection \ref{subsect:AnalysisDS2DKPIs} analyzes key RAN performance indicators, including RSRP, RSRQ, and SINR. Lastly, based on these metrics, in Subsection~\ref{subsect:performance_assessment} we estimate the expected capacity of DS2D data services, and discuss and quantify alternatives to improve it.

\subsection{Network Expansion vs.~Constellation Deployment} \label{subsection:observed_network_expansion}

\begin{figure}[!tb]
    \centering
    \includegraphics[width=\linewidth]{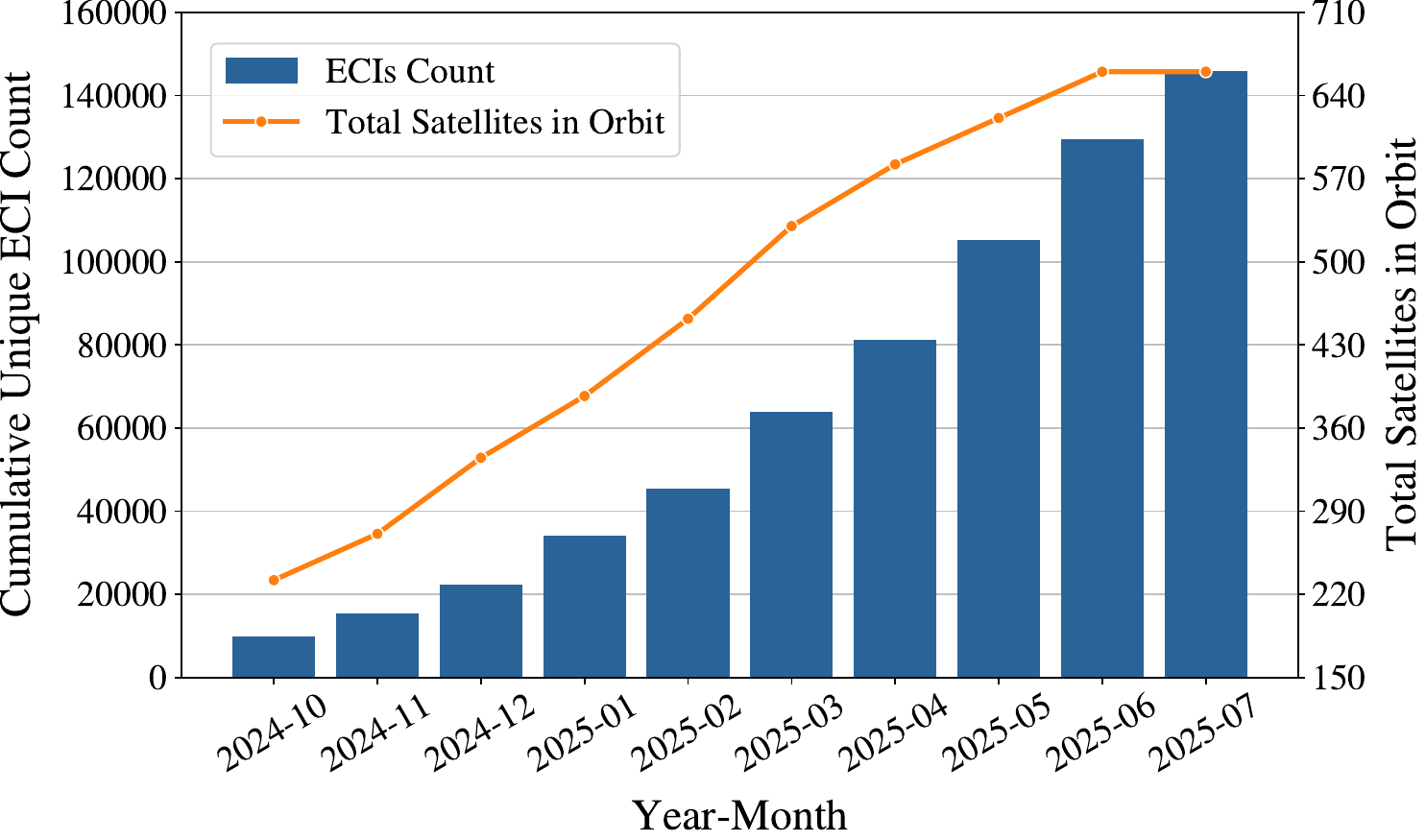}
    \caption{No.~monthly unique ECI vs.~No.~operational satellites.}
    \label{fig:uniqueECIs_satellites}
\end{figure}

Fig.~\ref{fig:uniqueECIs_satellites} shows the evolution of the number of cumulative unique cell identifiers (ECIs) associated with Starlink’s DS2D network, as observed in our dataset over the measurement period. The results reveal a steady upward trend, closely correlated with the timeline of Gen2 satellite launches. This finding suggests that crowdsourcing may serve as a useful tool for large-scale characterization of DS2D networks, and challenges the common perception that such data are inherently biased toward urban areas. On the contrary, our results demonstrate that crowdsourced measurements offer valuable insights in sparsely populated regions, which lack robust terrestrial coverage and are precisely where SCS services are expected to make the greatest impact.

\subsection{Starlink's DS2D Spatial Analysis} \label{subsec:Spatial_Coverage}

Although satellite constellations are designed to provide global coverage by definition, the potential use of SCS services is not geographically uniform. In practice, DS2D observability varies significantly across regions, influenced by factors such as terrestrial network availability, operator deployment strategies, population density, and accessibility.
\begin{figure*}
    \centering
    \includegraphics[width=\linewidth]{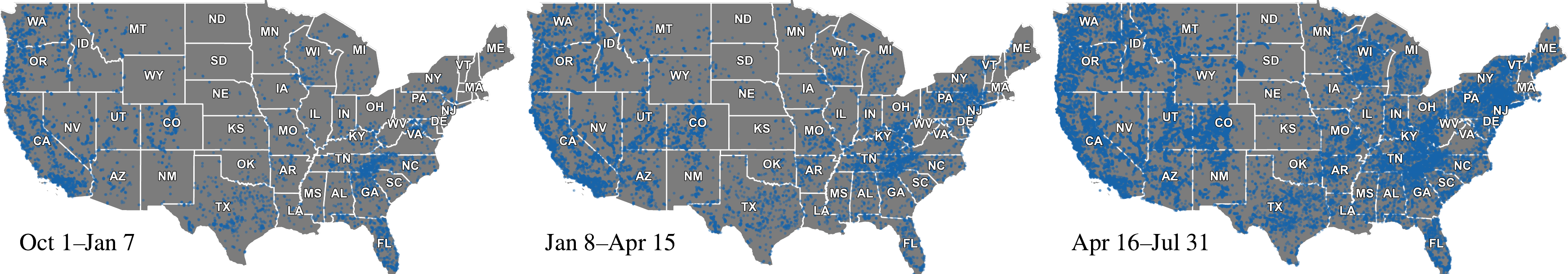}
    \caption{Spatio-temporal evolution of observed Starlink's DS2D measurements from October 2024 to July 2025.}
    \label{fig:4images}
\end{figure*}

Fig.~\ref{fig:4images} shows the evolution of the spatial distribution of Starlink's DS2D mobile crowdsourced measurements. Initial measurements in late 2024 were concentrated in southeastern states affected by hurricanes, where the service temporarily substituted inoperative terrestrial networks, with a comparable pattern observed during the January 2025 Southern California wildfires. Following the official beta launch in early 2025, measurements expanded nationwide, and by March--April 2025, adoption had intensified considerably. The measurements further suggest that service activation during the initial phases was geographically constrained until the constellation reached completion, as early activity was limited to specific regions.

\begin{figure}[!t]
    \centering
    \includegraphics[width=\linewidth]{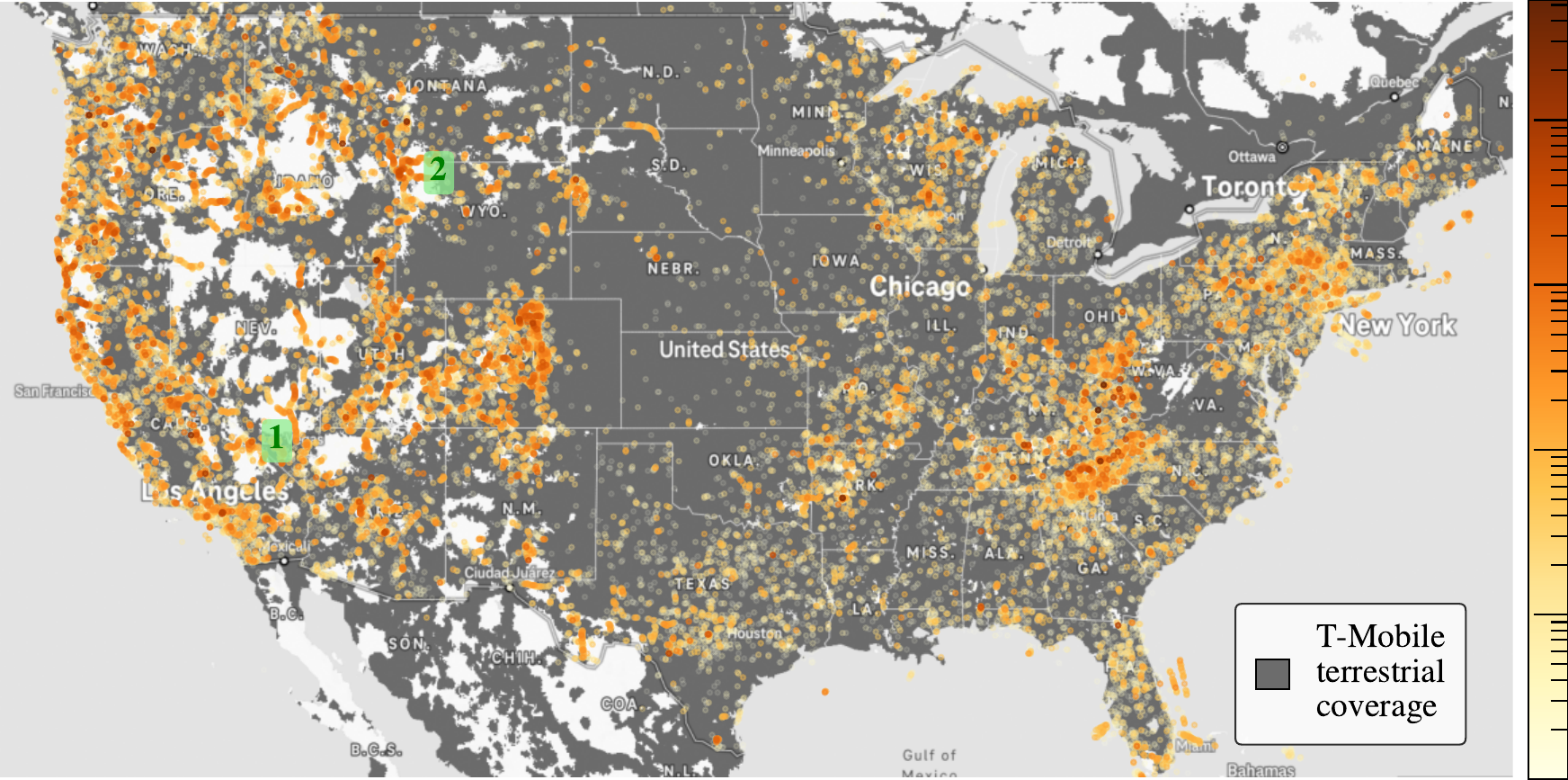}
    \caption{
    Starlink's measurement counts vs.~T-Mobile's coverage. 1. Death Valley National Park, 2. Yellowstone National Park.}
    \label{fig:T-Mobile_coverage}
\end{figure}

In Fig.~\ref{fig:T-Mobile_coverage}, we present a logarithmic-scale heatmap showing the number of observed Starlink DS2D measurements overlaid on T-Mobile’s public coverage map. 
The comparison offers several noteworthy insights. First, DS2D measurements are concentrated primarily in areas designated by T-Mobile as having satellite-only coverage, with some also appearing in areas nominally served by terrestrial networks, suggesting potential reliability issues. Second, most DS2D measurements are concentrated in accessible yet underserved regions near populated zones, particularly around national parks.  

\begin{figure}[!t]
    \centering
    \includegraphics[width=\linewidth]{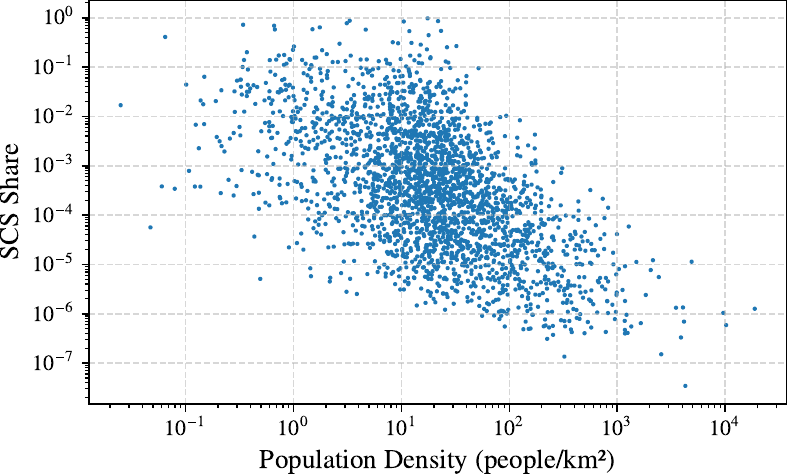}
    \caption{SCS share vs.~population density by county.}
    \label{fig:IntensityOfUse_vs_PopDensity}
\end{figure}

A quantitative analysis using the SCS share indicator (as defined in Section~\ref{Subsection:EstimatingSCSUse}) confirms that large national parks such as Death Valley and Yellowstone exhibit high reliance on DS2D, with values of $0.86$ and $0.80$, respectively. Figure~\ref{fig:IntensityOfUse_vs_PopDensity} plots the SCS share against population density, computed at a county level. Counties such as Knott (KY).  Florence (WI), and Martin (KY), also located near national parks, show SCS share values exceeding $0.85$. Overall, we find a moderately strong negative logarithmic correlation with county-level population density (Pearson correlation coefficient: $-0.57$).

\subsection{Analysis of DS2D Key Performance Indicators}
\label{subsect:AnalysisDS2DKPIs}

\begin{figure}[!t]
    \centering
    \includegraphics[width=\linewidth]{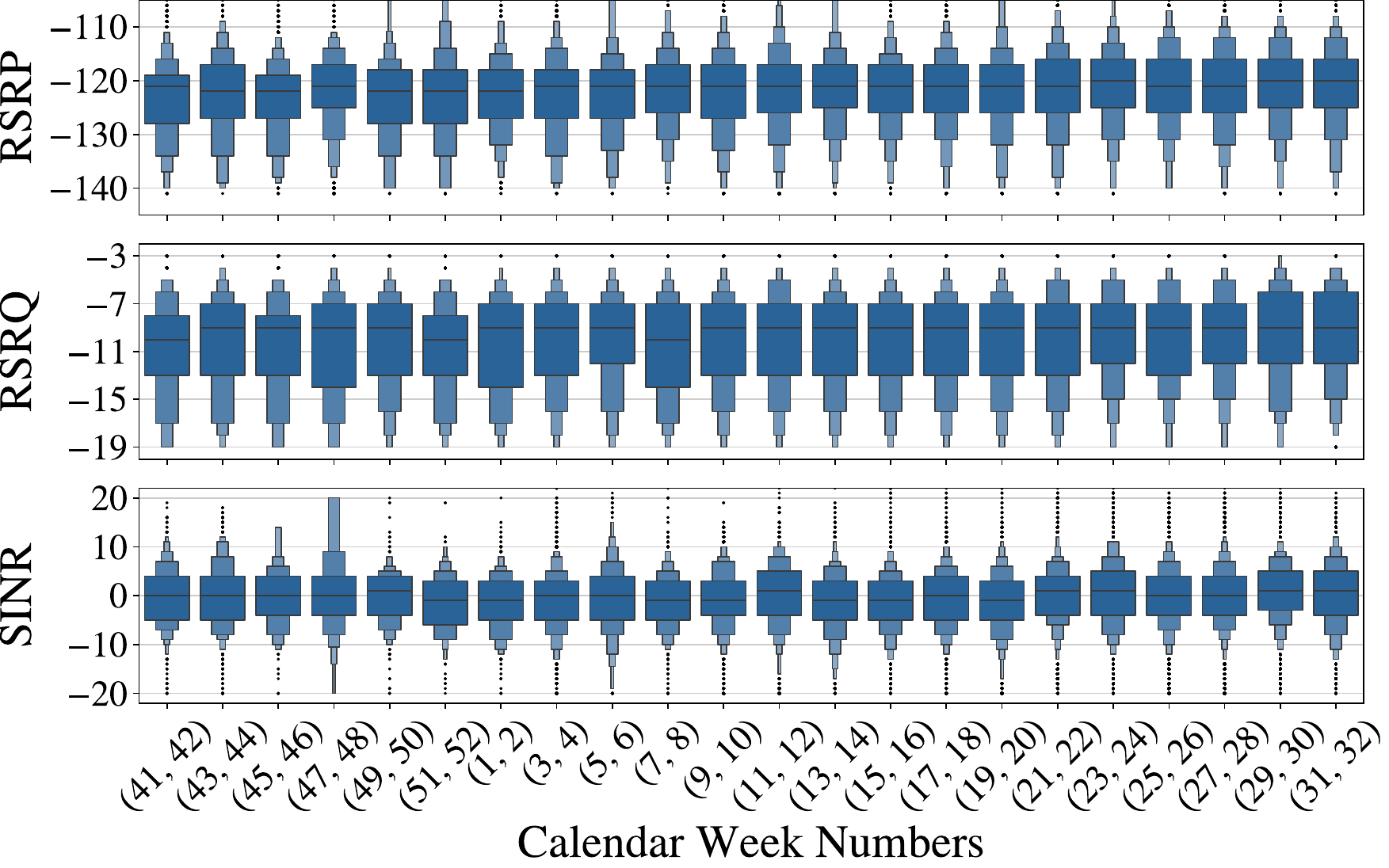}
    \caption{Biweekly letter-value plots of key RAN metrics.}
    \label{fig:LVPlot_RSRP_RSRQ}
\end{figure}

In this Subsection, we present key metrics of Starlink's DS2D RAN, namely RSRP, RSRQ and SINR. We examined the dataset for spatial patterns but did not identify any significant trends. In the temporal domain, we observed slight changes. Fig.~\ref{fig:LVPlot_RSRP_RSRQ} shows biweekly distributions (in the form of letter-value plots) of RSRP, RSRQ and SINR measurements from the first week of October 2024 (week 41) to the last week of July 2025 (week 32). This timeline captures key milestones of the service setup, including the authorization granted in March 2025 to increase OOBE by $9.4$ dB. The latter, due to transmitter non-linearities, would allow for an approximately $3.1$-dB increase in transmit power according to SpaceX declarations.\footnote{See David Goldman's (SpaceX) letter in response to Merissa Velez (FCC) on September 13, 2024. Electronic Comment Filing System (ECFS) code 10914226871971/1} A detailed analysis shows a slight improvement in the distribution of the RAN metrics over the measurement period, although the observed increment, in particular in RSRP, is smaller than the expected value.
 
\begin{figure*}[htbp]
    \centering
    \includegraphics[width=\linewidth]{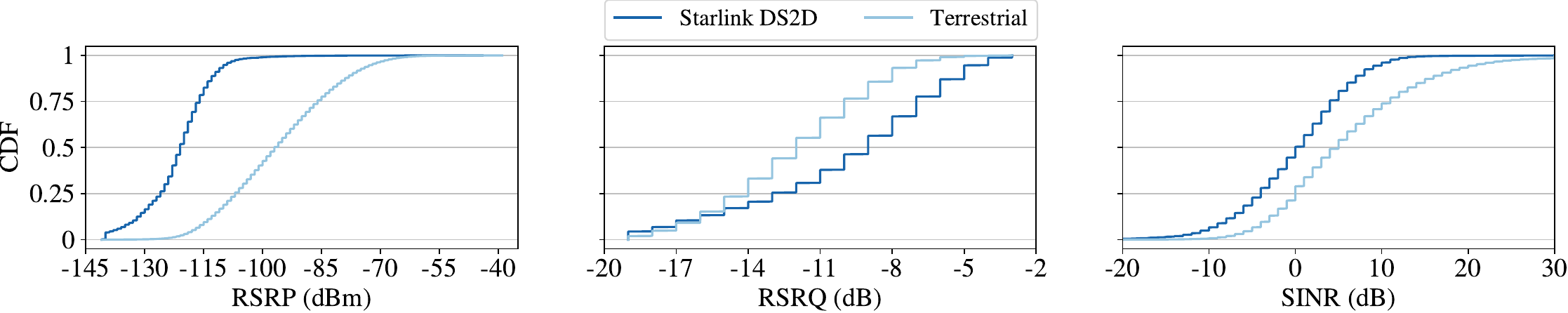}
    \caption{CDF of physical-layer measurements for Starlink DSD and T-Mobile's terrestrial networks.}
    \label{fig:Comparison_DS2D_Terrestrial}
\end{figure*}

A key question is how this DS2D RAN behavior differs from that of terrestrial cellular technologies. For this, we benchmark the metrics of this network with those of T-Mobile's LTE terrestrial network. Fig.~\ref{fig:Comparison_DS2D_Terrestrial} compares the cumulative distribution functions (CDF) of RSRP, RSRQ and SINR in Starlink's DS2D network and T-Mobile terrestrial network. DS2D shows substantially lower and more homogeneous RSRP levels (median: $–121$ dBm vs.~$-97$ dBm; interdecile range: $21$ dB vs.~$37$ dB), reflecting larger propagation losses and uniform link geometry, as users are located at large but relatively similar distances from the base stations, predominantly with line-of-sight conditions. Despite limited bandwidth, RSRQ is more favorable in DS2D (median: $-9$ dB vs.~$-12$ dB) owing to low network load and minimal interference, though this advantage may diminish as demand grows. The correlation between RSRP and RSRQ is stronger in DS2D ($0.69$ vs.~$0.34$), consistent with lower interference. SINR values are also lower in DS2D (median: $0$ dB vs.~$5$ dB), underscoring its coverage-limited, low-load operation.

\subsection{Performance assessment of future DS2D services}
\label{subsect:performance_assessment}

Building on the previous observations, we evaluate the potential performance of future DS2D data services. For this, we apply the modified Shannon capacity formulation (see Section \ref{Subsection:methodology_performance}) using the available bandwidth and SINR measurements of the last three months, to ensure that the results reflect a more stable operational state of the network. Based on the SINR distributions, the mean and median spectral efficiency values are $0.61$ and $0.52$ bps/Hz, respectively. Note that, given that high SINR values are rarely observed, this estimation is largely insensitive to the method used to compute $m$, and the influence of this parameter is negligible. 

Since Starlink’s current DS2D operations are restricted to the PCS G Block ($2 \times 5$ MHz), the derived spectral efficiency corresponds to an approximate downlink capacity of $3.1$ Mbps per beam. This value should be regarded as an upper-bound estimate of the average throughput per connection, corresponding to the scenario of a single user occupying the full bandwidth of the beam. While such speeds may be sufficient to provide basic connectivity in remote or underserved regions, they remain well below the average downlink rates offered by terrestrial networks. This gap highlights the complementary nature of DS2D services in their current phase, consistent with their role as an SCS solution.

Nevertheless, performance may improve in the coming months through several potential capacity expansion options, including (\textit{i}) higher transmit power, (\textit{ii}) reduced orbital altitude, (\textit{iii}) expanded bandwidth, and (\textit{iv}) the continued growth of the satellite constellation.

First, as discussed in Subsection~\ref{subsect:AnalysisDS2DKPIs}, we do not observe the $3.1$-dB increase in the RAN metrics that would be expected from applying the FCC's March 2025 authorization; instead, only a slight improvement is reflected in the measurements. With the available evidence, it is not possible to determine whether this variation is attributable to the regulatory change. It therefore remains unclear whether transmissions are already occurring at the newly authorized power levels, although SpaceX’s filings suggest that this is the case (see footnote 4). Nevertheless, in the hypothetical assumption that the $9.4$-dB relaxation in the OOBE limit allows for a future increase in transmit power, and assuming that the increase in SINR would be similar (coverage-limited network), we estimate that the spectral efficiency would increase from the current baseline to $0.89$ bps/Hz (mean) and $0.83$ bps/Hz (median), raising the estimated average per-beam capacity from $3.1$ Mbps to $4.5$ Mbps. 

Second, system performance may also be improved by lowering orbital altitudes. SpaceX has been authorized to selectively operate DS2D satellites in Very Low Earth Orbits (VLEO) at an altitude of $340$--$360$ km, subject to previous coordination with NASA.\footnote{See FCC’s DA 24-1193} Assuming free-space propagation, relocating the entire constellation to lower altitudes would reduce path loss by $20 \log_{10}(550/355) \approx 3.8$ dB. To comply with the OOBE constraints, this would require an estimated $1.3$-dB reduction in transmit power, resulting in a net gain of about $2.5$ dB in received signal power. Assuming again that the SINR values are increased in the same proportion, the spectral efficiency would raise from its baseline value to $0.83$ bps/Hz (mean) and $0.76$ bps/Hz (median). This corresponds to an approximate $1.4\times$ gain, yielding an estimated downlink capacity of $4.2$ Mbps per beam.

Third, system throughput may be increased by expanding the available bandwidth. In September 2025, Starlink acquired spectrum from EchoStar in the MSS AWS-4 band ($2000$--$2020$ MHz uplink, $2180$--$2200$ MHz downlink), and in the IMT PCS band (H Block, $1915$--$1920$ MHz uplink, $1995$--$2000$ MHz downlink), as discussed in Subsection~\ref{AWS-4, PCS H}. This additional spectrum provides significantly more capacity than the $2 \times 5$ MHz PCS G Block currently used for DS2D. 

Whereas the AWS-4 band has been standardized in 3GPP NTN as n252 and is expected to be integrated into smartphone chipsets in the future, the PCS H Block corresponds to existing IMT spectrum and can therefore be used immediately. Integrating the PCS H Block spectrum into the currently deployed infrastructure---which operates in the PCS G Block---would immediately double system capacity, increasing the per-beam estimate from $3.1$ Mbps to $6.2$ Mbps. In the medium term, once NTN bands are more widely supported in commercial devices, the additional $20$ MHz downlink from AWS-4 could be exploited. Combined with the PCS allocations, this would increase per-beam capacity by a factor of six relative to the baseline, reaching about $18.6$ Mbps when using both Starlink’s spectrum holdings and T-Mobile’s PCS G Block.

Beyond the per-beam capacity expected, these acquisitions reduce Starlink’s reliance on third-party terrestrial operators for spectrum access. By leveraging their own PCS H Block spectrum, Starlink can provide a service comparable to that currently offered through T-Mobile’s spectrum, with the estimated per-beam throughput of 3.1 Mbps. Furthermore, the aggregation of their full spectrum holdings in the AWS and PCS bands (totaling $2 \times 25$ MHz) would enable an estimated per-beam throughput of 15.5 Mbps, thereby allowing the provision of DS2D services without reliance on third-party spectrum agreements. Note that these figures correspond to aggregate per-beam capacity rather than achievable single-user rates. Attaining the maximum values per connection would require device support for spectrum aggregation across these multiple bands, which is not yet included in current NTN standardization. Accordingly, the realistic upper bound for per-user capacity is expected to be around $12.4$ Mbps (estimated throughput over the AWS band only).

In the longer term, satellite swarm antenna arrays show promising preliminary results in increasing the directivity of satellite beams~\cite{Tuzi23SatelliteSwarm}, which allows for higher gains and improved spatial multiplexing. Achieving this would require the launch and operation of many more satellites, which might also contribute to increasing the overall throughput of DS2D networks and the effective capacity experienced by end users. Although the FCC denied Starlink’s request to authorize $30\,000$ DS2D-capable satellites, the company remains authorized to operate up to $7\,500$ such satellites, indicating substantial potential for network densification. Furthermore, the aforementioned advancements in network capacity could be combined, potentially enabling greater per-beam throughput in the future.

\section{Conclusions} 
\label{Section:Conclusions}
In this article, we presented the first empirical analysis of a commercially deployed DS2D network, focusing on Starlink’s SCS Service in the U.S.~in partnership with T-Mobile. Using large-scale crowdsourced measurements between October 2024 and July 2025, we characterized the spatio-temporal evolution and RAN performance of the DS2D network, which only supported SMS during the observed period.

Our results confirm a strong correlation between the number of DS2D-capable satellites launched and the number of unique cell identifiers observed, as well as a spatial distribution of measurements consistent with the operator’s coverage gaps. We observe a significant number of measurements concentrated around areas that are relatively accessible yet underserved by terrestrial networks, such as national parks. Furthermore, we find significant negative correlation between the prevalence of Starlink measurements and population density, reflecting its role as a complementary solution in sparsely populated regions.

The measurements indicate slight increases in RSRP, RSRQ, and SINR throughout the beta testing period, which may be attributable to network changes, including the higher OOBE levels authorized by the FCC and the deployment of some satellites at VLEO altitudes. Based on the SINR measurements, we estimate the current average per-beam throughput to be approximately $3.1$ Mbps, which is sufficient for basic services but significantly below terrestrial averages. Looking ahead, per-beam throughput could increase to $4.2$ Mbps by using VLEO satellites ($\sim 350$ km), and up to $4.5$ Mbps for LEO, assuming that OOBEs relaxation has not been applied yet. In the short term, recent spectrum acquisitions will enable doubling per-beam capacity (to $6.2$ Mbps), with the potential to reach $18.6$ Mbps in the medium term once NTN bands are integrated into commercial smartphone modems. However, the extent to which these enhancements translate into higher per-user throughput will ultimately depend on the expansion of the satellite constellation and the evolution of service demand. 

Our work illustrates the value of crowdsourced data to monitor emerging network technologies. Beyond characterizing deployment and performance, such data can help detect underserved zones, assess coverage gaps, and inform infrastructure planning, offering a promising tool for both technical and policy research on mobile communication systems. Future work will involve monitoring the evolution of DS2D networks and services, tracking the evolution of Starlink’s network as new services are introduced, and exploring other emerging DS2D deployments.

\section*{Acknowledgments}

We would like to thank Dr.~Tim Farrar for bringing one of the references to our attention, and Dr.~Fadi Saibi for several corrections to an early version of the manuscript.

\bibliographystyle{IEEEtran}
\bibliography{references,local_references}

\section{Biography Section}

\vspace{-33pt}
\begin{IEEEbiographynophoto}{Jorge Garcia-Cabeza} is an Industrial Ph.D.~candidate at Univ.~Politécnica de Madrid (UPM) and Weplan Analytics. He holds MSc.~degrees in Telecommunications Engineering and Data Science from UPM. His research focuses on network measurements to characterize wireless technologies and spectrum utilization using crowdsourced data as a diagnostic tool.
\end{IEEEbiographynophoto}
\vspace{-33pt} 
\begin{IEEEbiographynophoto}{Javier Albert-Smet} is a Ph.D.~candidate at UPM focusing on mobile network measurements. He holds an MSc.~in Electrical Engineering from KTH, where he graduated with honors in 2022. His main research interests are AI \& mobile network optimization, a field in which he co-authored two patents at Ericsson.
\end{IEEEbiographynophoto}
\vspace{-33pt} 
\begin{IEEEbiographynophoto}{Zoraida Frias} is an Associate Professor at UPM, where she earned her Ph.D.~in Communications Technologies in 2016 (cum laude). Her work bridges engineering and policy, focusing on data-driven models for analyzing next-generation network deployments and their economic and regulatory dimensions, with a particular emphasis on spectrum.
\end{IEEEbiographynophoto}
\vspace{-33pt}
\begin{IEEEbiographynophoto}{Luis Mendo} is an Associate Professor at UPM. He received the Ph.D.~degree (cum laude) in Telecommunication Engineering from UPM in 2001. He has coauthored four books, four patents, and over $40$ papers in the area of mobile communications. His research interests include analysis and optimization of mobile networks, simulation methods, and statistical estimation.
\end{IEEEbiographynophoto}
\vspace{-33pt} 
\begin{IEEEbiographynophoto}{Santiago Andr\'es Azcoitia} is an Assistant Professor at UPM. He received the Ph.D.~in Telematics Engineering from Univ.~Carlos III de Madrid in 2023 (cum laude), and worked for $25$ years as an ICT strategy consultant. His research interests include telecommunication networks, digital economics, policy, regulations, and ethics.
\end{IEEEbiographynophoto}
\vspace{-33pt} 
\begin{IEEEbiographynophoto}{Eduardo Yraola} is the Lead Data Scientist of Weplan Analytics. He received his Ph.D.~in physics from Univ.~Autónoma de Madrid in 2015, and has over 10 years of experience as a data scientist.
\end{IEEEbiographynophoto}
\end{document}